\documentclass[12pt,onecolumn,draftcls]{IEEEtran}
\usepackage{graphicx,amssymb,amsmath}
\usepackage{multicol}
\usepackage[noadjust]{cite}
\usepackage{setspace}
\usepackage{stfloats}
\usepackage{midfloat}
\usepackage[normal]{threeparttable}
\usepackage{flushend,cuted}
\usepackage{bm}
\usepackage{textcomp}
\usepackage{latexsym,bm}
\usepackage{booktabs}
\usepackage{xcolor}
\usepackage{changebar}
\usepackage{graphicx}
\usepackage{subfigure}
\newtheorem{lemm}{Lemma}

\IEEEoverridecommandlockouts
\begin{document}
\title{New Results on the Fluctuating Two-Ray Model with Arbitrary Fading Parameters and Its Applications}
\author{Jiayi~Zhang,~\IEEEmembership{Member,~IEEE,} Wen Zeng, Xu Li, Qiang Sun, and Kostas P. Peppas,~\IEEEmembership{Senior Member,~IEEE}

\thanks{Copyright (c) 2015 IEEE. Personal use of this material is permitted. However, permission to use this material for any other purposes must be obtained from the IEEE by sending a request to pubs-permissions@ieee.org.}
\thanks{This work was supported in part by the National Natural Science Foundation of China (Grant No. 61601020), the Fundamental Research Funds for the Central Universities (Grant Nos. 2016RC013, 2017JBM319, and 2016JBZ003), and the Open Research Fund of National Mobile Communications Research Laboratory, Southeast University (No. 2015D02).}
\thanks{J. Zhang, W. Zeng, and X. Li are with the School of Electronic and
Information Engineering, Beijing Jiaotong University, Beijing 100044, P. R.
China (e-mail: jiayizhang@bjtu.edu.cn).}
\thanks{Q. Sun is with School of Electronic and information, Nantong
University, Nantong 226019, P. R. China. He is also with National Mobile Communications Research Laboratory, Southeast University, Nanjing 210096. }
\thanks{K. P. Peppas is the Department of Informatics and Telecommunications,
University of Peloponnese, 22100 Tripoli, Greece (e-mail: peppas@uop.gr).}
}
\maketitle
\begin{abstract}
The fluctuating two-ray (FTR) fading model provides a much better fit than other fading models for small-scale fading measurements in millimeter wave communications. In this paper, using a mixture of gamma distributions, new exact analytical expressions for the probability density and cumulative distribution functions of the FTR distribution with arbitrary fading parameters are presented. Moreover, the performance of digital communication systems over the FTR fading channel is evaluated in terms of the channel capacity and the bit error rate. The interaction between channel fading  parameters and system performance is further investigated. Our newly derived results extend and complement previous knowledge of the FTR fading model.
\end{abstract}
\IEEEpeerreviewmaketitle

\section{Introduction}
One of the major challenges for future generation wireless
networks is the efficient utilization of available spectrum resources. The millimeter wave (mmWave) and device-to-device (D2D), are regarded as promising technologies to this end \cite{Boccardi2013Five,wong2017key}. Further research effort has focused on the characterization of the mmWave and D2D channel, e.g. see \cite{Rappaport2015Wideband, Peng2014Device, Cotton2015Human, Yoo2016Shadowed}.
Recently, the so-called fluctuating two-ray (FTR) fading model, has been proposed as a versatile model that well characterizes wireless propagation
in mmWave and D2D environments \cite{Romero2016The}. In contrast to the two-wave with diffuse power (TWDP) fading model \cite{Rao2015MGF}, the specular components of the FTR model are varying  amplitudes rather than constant amplitudes, which bring a better description of amplitude fluctuations. FTR fading model
fits well experimental channel characterization/modeling data, such as those obtained by outdoor millimeter-wave field measurements at 28 GHz \cite{Samimi2016mili},
and includes several well known distributions, i.e. the Gaussian, Rayleigh, Rician and Nakagami-$m$ ones, as special or limiting cases \cite{Romero2016The}.

The probability density function (PDF) and cumulative distribution function (CDF) of the FTR fading model, assuming integer values of its shadowing parameter, $m$, have been derived in \cite{Romero2016The} by employing inverse Laplace transforms. In the same work, approximate expressions for the PDF and CDF have also been presented in terms of a mixture of gamma distributions, under the assumption of integer $m$. In realistic propagation scenarios, however, $m$ is an arbitrary positive real number.
Motivated by the above facts, this paper extends the work of \cite{Romero2016The} and derives a generic analytical framework for the statistical characterization of the FTR fading model, assuming arbitrary positive values of $m$, and in terms of elementary functions and coefficients consisting of fading parameters. Also, our derived results facilitate the performance analysis of wireless communication systems operating in the FTR fading channel. 
Based on the above formula, novel analytical expressions for the channel capacity of FTR fading channels and the bit error rate (BER) of various binary modulation formats are deduced. Asymptotic BER results that become tight at high-SNR values are further presented, that offer valuable insights as to the impact of fading parameters on the system performance. The results presented herein enable the evaluation of critical performance metrics at low computational complexity, and thus, they are useful to the system engineer for performance evaluation purposes.

\section{Statistical Characterization Of The FTR Fading Model}\label{se:pdf and cdf}
In this section, the statistical properties of the FTR fading model are investigated. To this end, novel, analytical expressions for its PDF and CDF are derived for arbitrary values of its parameters.

\subsection{An Overview of the FTR Fading Model}\label{se:model}
According to the FTR fading model, the complex baseband response of the wireless channel can be expressed as
\begin{equation}\label{Vr2}
{V_r} = \sqrt \zeta  {V_1}\exp \left( {\imath{\phi _1}} \right) + \sqrt \zeta  {V_2}\exp \left( {\imath{\phi _2}} \right) + X + \imath Y,
\end{equation}
where $\zeta $ is a Gamma distributed random variable with unit mean, and PDF given by
\begin{equation}\label{fu}
{f_\zeta }\left( u \right) = \frac{{{m^m}{u^{m - 1}}}}{{\Gamma \left( m \right)}}{e^{ - mu}.}
\end{equation}
Furthermore, ${{V_1}}$ and ${{V_2}}$ are constant amplitudes having specular components modulated by a Nakagami-$m$ random variable, $\imath$ denotes the imaginary term, ${\phi _1}$ and ${\phi _2}$ are uniformly distributed random phases, namely ${\phi _1}, {\phi _2} \sim \mathcal{U}\left[ {0,2\pi } \right)$. The random phase of each dominant wave is assumed to be statistically independent. In addition, ${X + {\imath}Y}$ represents the diffuse component, which can be modeled as a complex Gaussian random variable as $X,Y \sim \mathcal{N}(0,{\sigma ^2})$.

The FTR fading model can be conveniently expressed by introducing the parameters $K$ and $\Delta $, which are respectively defined as
\begin{equation}\label{K}
K = \frac{{V_1^2 + V_2^2}}{{2{\sigma ^2}}},
\end{equation}
\begin{equation}\label{deta}
\Delta {\text{ = }}\frac{{2{V_1}{V_2}}}{{V_1^2 + V_2^2}},
\end{equation}
where $K$ is the ratio of the average power of the dominant waves to the average power of the remaining
diffuse multipath, in a similar fashion as the parameter $K$ in the Rician channel model. Moreover, $\Delta $ varies from 0 to 1 and characterizes the relation between the powers of two dominant waves. When $\Delta = 1$, the magnitudes of the two specular components are equal, e.g., ${V_1} = {V_2}$.  When $\Delta = 0$, the FTR model reduces to a Rician shadowed fading model, including only one component, e.g., ${V_1} = 0$ or ${V_2} = 0$. By using the definitions for $m$, $K$ and $\Delta $, the FTR fading model encompasses the one-sided Gaussian, Rayleigh, Nakagami-$q$, Rician, Rician shadowed and TWDP, as special cases.
\subsection{New expressions for the PDF and CDF of the FTR fading model}
The received average SNR, ${\bar \gamma }$, undergoing a multipath fading channel as described in (\ref{Vr2}) is given by
\begin{align}\label{averSNR}
\bar \gamma   = \left( {{{{E_b}} \mathord{\left/
 {\vphantom {{{E_b}} {{N_0}}}} \right.
 \kern-\nulldelimiterspace} {{N_0}}}} \right){\rm \mathbb{E}}\left\{ {{{\left| {{V_r}} \right|}^2}} \right\}
  = \left( {{{{E_b}} \mathord{\left/
 {\vphantom {{{E_b}} {{N_0}}}} \right.
 \kern-\nulldelimiterspace} {{N_0}}}} \right)2{\sigma ^2}\left( {1 + K} \right),
\end{align}
where ${E_b}$ is the energy density and ${\rm \mathbb{E}}\left\{  \cdot  \right\}$ denotes the expectation operator.

The PDF expression for the FTR fading power envelope can be obtained by ${K_u} = uK$ over all possible realizations $u$ of the random variable $\zeta $, which follows a Gamma distribution as indicated in (\ref{fu}). It should be stressed that the PDF and CDF expressions presented in \cite{Romero2016The}, are only valid for positive integer values of $m$. In what follows, novel PDF and CDF expressions for the power envelope of the FTR fading model are derived, assuming arbitrary positive real values of $m$.

\begin{lemm}
For arbitrary positive values of $m$, the exact PDF and CDF of the instantaneous SNR in the FTR fading channel can be respectively expressed as
\begin{equation}\label{PDF}
{f_\gamma }\left( x \right) = \frac{{{m^m}}}{{\Gamma \left( m \right)}}\sum\limits_{j = 0}^\infty  {\frac{{{K^j}{d_j}}}{{j!}}{f_G}\left( {x;j + 1,2{\sigma ^2}} \right)},
\end{equation}
\begin{equation}\label{CDF}
{F_\gamma }\left( x \right) = \frac{{{m^m}}}{{\Gamma \left( m \right)}}\sum\limits_{j = 0}^\infty  {\frac{{{K^j}{d_j}}}{{j!}}{F_G}\left( {x;j + 1,2{\sigma ^2}} \right)},
\end{equation}
where
\begin{equation}\label{fg}
{f_G}\left( {x;j + 1,2{\sigma ^2}} \right) \triangleq \frac{{{x^j}}}{{\Gamma \left( {j + 1} \right){{\left( {2{\sigma ^2}} \right)}^{j + 1}}}}\exp ( - \frac{x}{{2{\sigma ^2}}}){\text{   }},
\end{equation}
\begin{equation}\label{FG}
{F_G}\left( {x;j + 1,2{\sigma ^2}} \right) \triangleq \frac{1}{{\Gamma \left( {j + 1} \right)}}\gamma \left( {j + 1,\frac{x}{{2{\sigma ^2}}}} \right),
\end{equation}
\begin{align}\label{dj}
  &{d_j} \triangleq\sum\limits_{k = 0}^j {\left( {\begin{array}{*{20}{c}}
  j \\
  k
\end{array}} \right){{\left( {\frac{\Delta }{2}} \right)}^k}\sum\limits_{l = 0}^k {\left( {\begin{array}{*{20}{c}}
  k \\
  l
\end{array}} \right)} } \Gamma \left( {j \!+\! m \!+\! 2l \!-\! k} \right){e^{\frac{{\pi (2l \!-\! k)i}}{2}}} \hfill \notag\\
  & \times {\left( {{{(m + K)}^2} \!-\! {{(K\Delta )}^2}} \right)^{\frac{{ -(j+m)}}{2}}}  P_{j \!+\! m \!-\! 1}^{k \!-\! 2l}\left( {\frac{{m \!+\! K}}{{\sqrt {{{(m \!+\! K)}^2} \!-\! {{(K\Delta )}^2}} }}} \right) .\notag
\end{align}
where $\gamma \left( { \cdot , \cdot } \right)$ is the incomplete gamma function \cite[Eq. (8.350.1)]{Gradshteyn1980In} and $P\left(  \cdot  \right)$ denotes Legendre functions of the first kind \cite[Eq. (8.702)]{Gradshteyn1980In}.
\end{lemm}
\begin{IEEEproof}
Please see Appendix.
\end{IEEEproof}

\begin{figure}[htbp]
\centering
\includegraphics[scale=0.5]{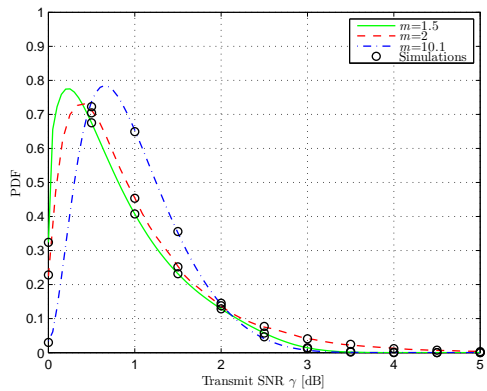}
\caption{FTR power envelope distribution for arbitrary positive values of $m$ ($K$ = 15, $\Delta$ = 0.5 and ${\bar \gamma }$ = 1). }
\label{pdfm}
\end{figure}

In contrast to \cite{Romero2016The}, the derived PDF and CDF expressions in Lemma 1 are more general and valid for arbitrary positive values of $m$. Fig. \ref{pdfm} depicts the PDF derived in (\ref{PDF})  for different arbitrary positive values of $m$, and markers correspond to the simulation PDF based on (\ref{Vr2}). It is clear that the Monte Carlo simulations validate our derived result of PDF, and the difference of the PDF curves between considering $m=1.5$ and $m=2$ is pronounced.

\subsection{KS Goodness-of-fit test}
Herein, we investigate the validity of the proposed approximations by using the statistical tool and argument. Specifically, we employ the Kolmogorov-Smirnov (KS) goodness-of-fit statistical test \cite{Papoulis1965Probability}, which measures the maximum value of the absolute difference between the empirical CDF of the random variable $\gamma $, ${{\hat F}_{ \gamma }}\left(  \cdot  \right)$,  and the analytical CDF of the random variable ${ \gamma }$, ${F_\gamma }\left(  \cdot  \right)$. Hence, the KS test statistic is given by \cite{Papoulis1965Probability}
\begin{equation}\label{KS}
T \triangleq \max \left|{{{\hat F}_\gamma }\left( x \right)} -{F_\gamma }\left( x \right) \right|.
\end{equation}

Table \ref{tab1} depicts the KS test statistic for different combinations of channel parameters $K$, $m$ and $\Delta$, where ${T_1}$ is for the proposed approximation in this paper, and ${T_2}$ is for the approximation given in \cite{Romero2016The}. It should be noted that, for all considered  cases in this paper, 40 terms have been used to converge the series. Without loss of generality, we assume that the average SNR ${\bar \gamma = 1}$. The exact results ${F_\gamma }\left( x \right)$ have been obtained by averaging at least $v = {10^4}$ samples of the FTR random variables. The critical value ${T_{\max }} = \sqrt { - \left( {{1 \mathord{\left/
 {\vphantom {1 {2v}}} \right.
 \kern-\nulldelimiterspace} {2v}}} \right)\ln \left( {{\alpha  \mathord{\left/
 {\vphantom {\alpha  2}} \right.
 \kern-\nulldelimiterspace} 2}} \right)} $, is given by ${T_{\max }} = 0.0136$, which corresponds to a significance level of $\alpha  = 5\% $ \cite{Papoulis1965Probability}. The hypothesis ${H_0}$ is accepted with $T < {T_{\max }}$. It is clearly illustrated in Table \ref{tab1} that hypothesis ${H_0}$ is always accepted with $95\% $ significance for different combinations of parameters $K$, $m$ and $\Delta$, as both ${T_1}$ and ${T_2}$ are smaller than ${T_{\max }}$. As opposed to \cite{Romero2016The}, our derived approximations are valid for arbitrary positive values of parameters $m$, which is practical in a real wireless scenario. More specifically, the hypothesis ${H_0}$ of the CDF expression (\ref{CDF}) is also accepted for arbitrary positive values of $m$. It should be noted that, for all cases considered in this paper, 40 terms have been used to converge the series and the truncation error is less than $10^{-9}$.

\begin{table}[!t]
\renewcommand{\thetable}{\Roman{table}}
\caption{ statistic KS test for different combinations of channel parameters $K$, $m$, $\Delta$.}
\label{tab1}
\newcommand{\tabincell}[2]{\begin{tabular}{@{}#1@{}}#2\end{tabular}}
\centering          
\begin{tabular}{|c|c|c|c|c|}        
\hline
\hline
FTR Fading Parameters & ${T_1}$ & ${T_2}$\\
\hline
\hline
 $m$=5.5, $K$=15, $\Delta$=0.4 &  0.013339 &  --  \\
\hline
 $m$=8.5, $K$=5,  $\Delta$=0.35 &  0.011112 &  --  \\
\hline
$m$=9.2, $K$=3,  $\Delta$=1 &  0.009313  &  -- \\
\hline
 $m$=10,  $K$=10, $\Delta$=0.5 &  0.008867 &  0.008867  \\
\hline
 $m$=15,  $K$=20, $\Delta$=0.2 &  0.005461 &  0.005461  \\
\hline
 $m$=20,  $K$=5,  $\Delta$=0.43 &  0.009404 &  0.009404  \\
\hline
\hline
\end{tabular}
\end{table}


%

\section{Performance Analysis}\label{se:uplink}

\subsection{Channel Capacity}
The average capacity per unit bandwidth is given by
\begin{equation}\label{Cora}
{C} \triangleq \int_0^\infty  {{{\log }_2}\left( {1 + x} \right){f_\gamma }\left( x \right)dx} .
\end{equation}
With the help of (\ref{Cora}) and (\ref{PDF}), we can obtain the capacity of the FTR channels.

\begin{lemm}
For arbitrary positive values of $m$, the average FTR channel capacity per unit bandwidth can be obtained as
\begin{equation}\label{Cftr}
{C_{{\text{\emph{FTR}}}}} = \frac{{{m^m}}}{{\Gamma \left( m \right)}}\sum\limits_{j = 0}^\infty  {\frac{{{K^j}{d_j}}}{{j!}}{L_G}\left( {j + 1,2{\sigma ^2}} \right)},
\end{equation}
where
\begin{equation}
{L_G}\left( {j \!+\! 1,2{\sigma ^2}} \right) \triangleq \frac{{\exp \left( \frac{1}{{2{\sigma ^2}}}  \right)}}{{\ln 2}}\sum\limits_{k = 0}^j {{\left( {{2{\sigma ^2}}}\right) ^{\!-\!k}}\Gamma \left( { \!-\! k,\frac{1}{{2{\sigma ^2}}} } \right)}.\notag
\end{equation}
\end{lemm}
\begin{IEEEproof}
Substituting (\ref{PDF}) into (\ref{Cora}), we can obtain
\begin{equation}
{C} = \frac{{{m^m}}}{{\Gamma \left( m \right)}}\sum\limits_{j = 0}^\infty  {\frac{{{K^j}{d_j}}}{{j!}}} \int_0^\infty  {{{\log }_2}\left( {1 + x} \right){f_G}\left( {x;j + 1,2{\sigma ^2}} \right)dx}\notag
\end{equation}
and we define
\begin{equation}
{L_G}\left( {j + 1,2{\sigma ^2}} \right) \triangleq \int_0^\infty  {{{\log }_2}\left( {1 + x} \right){f_G}\left( {x;j + 1,2{\sigma ^2}} \right)dx}.\notag
\end{equation}
The closed-form expression of ${L_G}\left( {j + 1,2{\sigma ^2}} \right)$ can be easily derived to finish the proof.
\end{IEEEproof}

\vspace{-3mm}
\subsection{Bit Error Rate}
For a variety of modulation formats, the average BER is given by
\begin{equation}\label{Pber}
{P_{{\text{BER}}}} \triangleq \int_0^\infty  {{P_e}\left( x \right){f_\gamma }\left( x \right)dx} ,
\end{equation}
where ${{P_e}\left( x \right)}$ is the conditional bit-error probability, which can be written as
\begin{equation}\label{Pex}
{P_e}\left( x \right) = \frac{{\Gamma \left( {\beta ,\alpha x} \right)}}{{2\Gamma \left( \beta  \right)}},
\end{equation}
where ${\Gamma \left( {\beta ,\alpha x} \right)}$ is the upper incomplete Gamma function \cite[Eq. (8.350.2)]{Gradshteyn1980In}, $\alpha$ and $\beta$ are modulation-specific parameters for binary modulation schemes, respectively. For example, $\left( {\alpha ,\beta } \right) = \left( {1,0.5} \right)$ for binary shift keying (BPSK), $\left( {\alpha ,\beta } \right) = \left( {0.5,0.5} \right)$ for coherent binary frequency shift keying, and $\left( {\alpha ,\beta } \right) = \left( {1,1} \right)$ for differential BPSK \cite{Trigui2009Performance}. By substituting (\ref{PDF}) and (\ref{Pex}) into (\ref{Pber}) and based on the definition of ${\Gamma \left( {\beta ,\alpha x} \right)}$ \cite{Zhang2012Performance}, (\ref{Pber}) can be rewritten as
\begin{equation}\label{Pber1}
{P_{{\text{BER}}}} = \frac{{{\alpha ^\beta }}}{{2\Gamma \left( \beta  \right)}}\int_0^\infty  {{x^{\beta  - 1}}{e^{ - \alpha x}}{F_\gamma }\left( x \right)dx}.
\end{equation}
With the help of (\ref{Pber1}) and (\ref{CDF}), we can obtain the BER of the FTR fading channel in the following lemma.

\begin{lemm}
For arbitrary positive values of $m$, the average BER of the FTR fading channel can be obtained as
\begin{equation}\label{Pberftr}
{P_{{\text{\emph{BER}}}}} = \frac{{{m^m}}}{{\Gamma \left( m \right)}}\sum\limits_{j = 0}^\infty  {\frac{{{K^j}{d_j}}}{{j!}}{B_G}\left( {j + 1,2{\sigma ^2}} \right)},
\end{equation}
where
\begin{align}\label{Bg}
&{B_G}\left( {j + 1,2{\sigma ^2}} \right)\triangleq \frac{{\Gamma \left( {\beta  + j + 1} \right){{\left( {2{\sigma ^2}} \right)}^\beta }}}{{\left( {j + 1} \right){{\left( {2\alpha {\sigma ^2} + 1} \right)}^{\beta  + j + 1}}}}\notag\\
  & \times {}_2{F_1}\left( {1,\beta  + j + 1;j + 2;\frac{1}{{1 + 2\alpha {\sigma ^2}}}} \right),
\end{align}
where ${}_2{F_1}\left( { \cdot , \cdot ; \cdot ; \cdot } \right)$ is the Gauss hypergeometric function \cite[Eq. (9.14)]{Gradshteyn1980In}.
\end{lemm}
\begin{IEEEproof}
Substituting (\ref{CDF}) into (\ref{Pber1}), we can obtain
\begin{align}\label{CDF2Pber1}
  {P_{{\text{BER}}}} &=  \frac{{{m^m}}}{{\Gamma \left( m \right)}}\sum\limits_{j = 0}^\infty  \frac{{{K^j}{d_j}}}{{j!\Gamma \left( {j + 1} \right)}}\frac{{{\alpha ^\beta }}}{{2\Gamma \left( \beta  \right)}}\notag \\
& \times \underbrace {\int_0^\infty  {{x^{\beta  - 1}}{e^{ - \alpha x}}\gamma \left( {j + 1,\frac{x}{{2{\sigma ^2}}}} \right)dx} }_{{z_j}}  \hfill
\end{align}
With the help of \cite[Eq. (6.455.2)]{Gradshteyn1980In}, the integral ${{z_j}}$ can be expressed as
\begin{align}\label{zj}
{z_j} &= \frac{{\Gamma \left( {\beta  + j + 1} \right){{\left( {2{\sigma ^2}} \right)}^\beta }}}{{\left( {j + 1} \right){{\left( {2{\sigma ^2}\alpha  + 1} \right)}^{\beta  + j + 1}}}} \notag  \\
& \times {}_2{F_1}\left( {1,\beta  + j + 1;j + 2;\frac{1}{{1 + 2\alpha {\sigma ^2}}}} \right).
\end{align}
The proof concludes by combining (\ref{zj}) and (\ref{CDF2Pber1}).
\end{IEEEproof}

Note that the derived BER expression \eqref{Pberftr} is given in terms of Gauss hypergeometric functions, which can be easily evaluated and efficiently programmed in most standard software packages (e.g., Matlab, Maple and Mathematica). However, our exact analytical results provide limited physical insights, we now present an asymptotic and simple expression of the error rates for the high-SNR regime.

For the high-SNR regime, ${2{\sigma ^2}}$ approaches to $\infty $, resulting in the term ${B_G}\left( {j + 1,2{\sigma ^2}} \right)$ approaches zero. Therefore, the term with $j = 0$ is the maximum and we can remove other terms. After some simple manipulations, we obtain the average BER in the high-SNR regime for arbitrary positive values of $m$ as
\begin{align}\label{PberftrH}
P_{{\text{\emph{BER}}}}^\infty  &= \frac{{{m^m}}}{{2\Gamma \left( \beta  \right)}}\frac{{\Gamma \left( {\beta  + 1} \right)}}{{2\alpha {\sigma ^2}}}{\left( {{{(m + K)}^2} - {{(K\Delta )}^2}} \right)^{\frac{{ - m}}{2}}}\notag\\
  & \times{P_{m - 1}}\left( {\frac{{m + K}}{{\sqrt {{{(m + K)}^2} - {{(K\Delta )}^2}} }}} \right).
\end{align}

Note that \eqref{PberftrH} is coincident with the expression in \cite{Romero2016The}, which is obtained from the asymptotic MGF for the high-SNR regime. Herein, we provide a method to obtain the BER for the high-SNR regime directly from the derived analytical BER expression.


\section{Numerical Results}\label{se:numerical_result}
In this section, some representative plots that illustrate the capacity and BER performance of FTR fading channels are presented, {along with the Monte Carlo simulation by generating ${10^8}$ random realizations following the FTR distribution.} Although some expressions presented herein are given in terms of infinite series, we only need use less than 40 terms to get a satisfactory accuracy (e.g., smaller than ${10^{ - 9}}$) for all considered cases. In the following, without loss of generality, the transmit SNR is normalized to be 1.

\begin{figure}[t]
\centering
\includegraphics[scale=0.5]{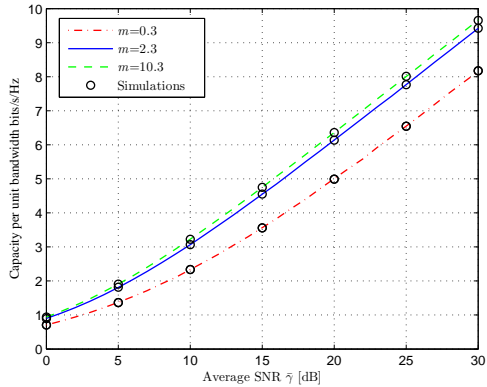}
\caption{Capacity under unit bandwidth of the FTR fading channel against the average SNR ${\bar \gamma }$ for different values of $m$ ($K$ = 10 and $\Delta$ = 0.5). }
\label{oram}
\end{figure}

\begin{figure}
\begin{minipage}[t]{0.5\linewidth}
\centering
\subfigure[$\Delta$ = 0.9]{
\includegraphics[width=1.6in,height=2.2in]{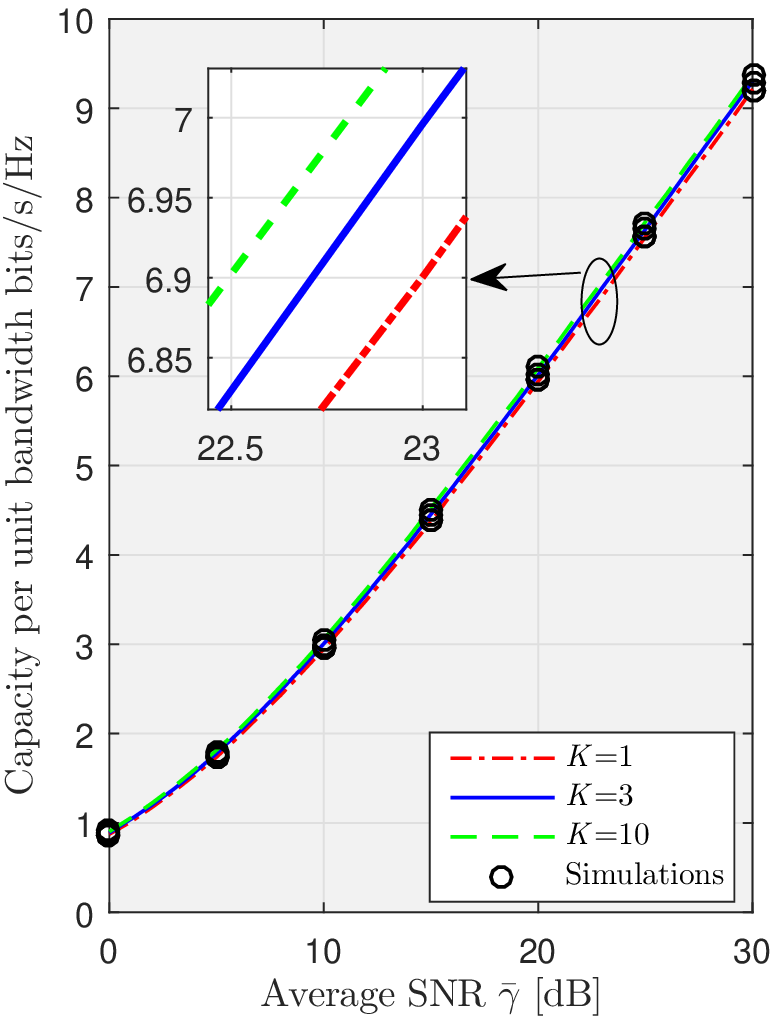}}
\label{fig:side:a}
\end{minipage}%
\begin{minipage}[t]{0.5\linewidth}
\centering
\subfigure[$\Delta$ = 1]{
\includegraphics[width=1.6in,height=2.2in]{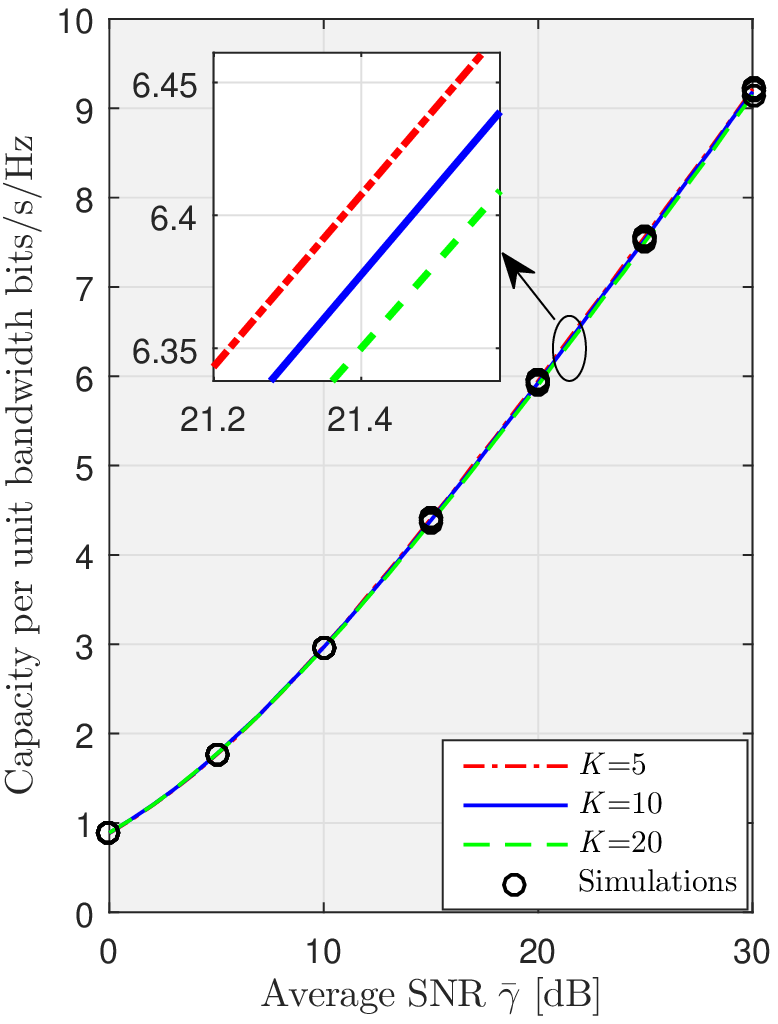}}
\label{fig:side:b}
\end{minipage}
\caption{Capacity under unit bandwidth of the FTR fading channel against the average SNR ${\bar \gamma }$ for different values of $K$ ($m$ = 25.5).}
\label{oraK}
\end{figure}

\begin{figure}[t]
\centering
\includegraphics[scale=0.5]{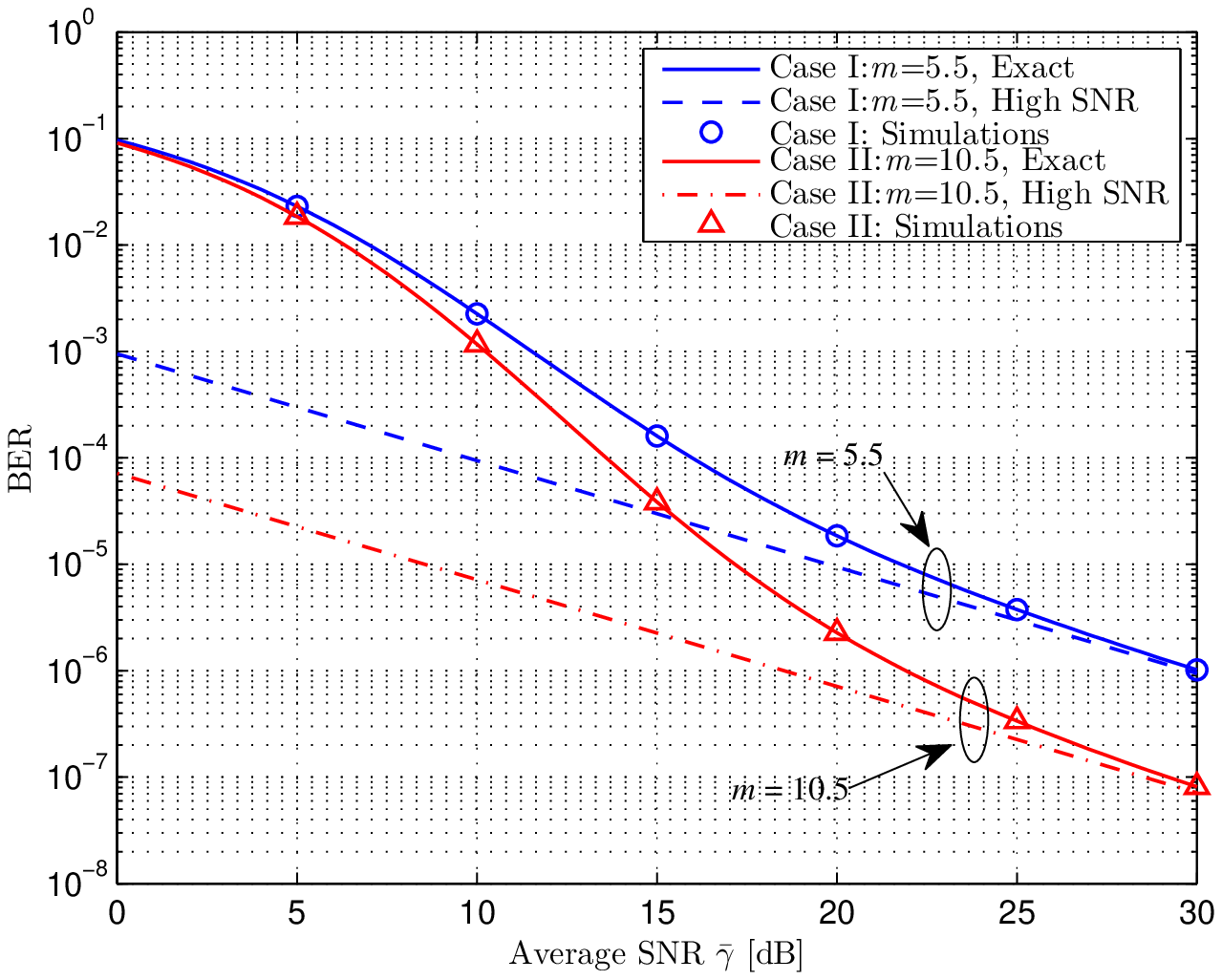}
\caption{BER performance of the FTR fading channel against the average SNR ${\bar \gamma }$ for different values of $m$ ($K$ = 30 and $\Delta$ = 0.45). }
\label{berm}
\end{figure}

\begin{figure}[t]
\centering
\includegraphics[scale=0.5]{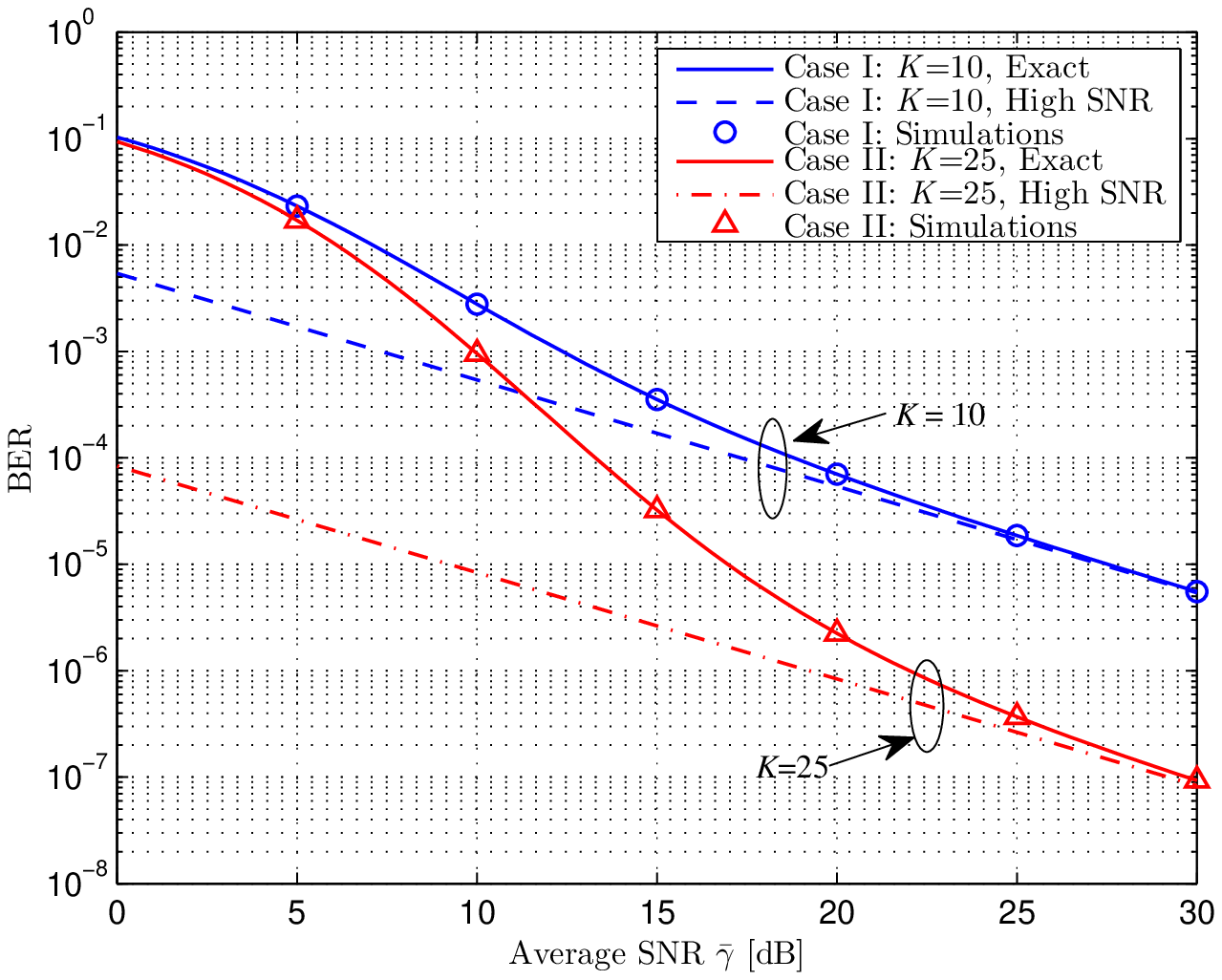}
\caption{BER performance of the FTR fading channel against the average SNR ${\bar \gamma }$ for different values of $K$ ($m$ = 10.5 and $\Delta$ = 0.35). }
\label{berK}
\end{figure}

{Figs. \ref{oram}-\ref{oraK} depict the analytical (\ref{Cftr}) and simulation channel capacity against the average SNR ${\bar \gamma }$ for different values of $m$ and $K$. By varying one parameter while keeping other parameters fixed, we find that increasing the values of $m$ helps overcome the effect of the FTR channel fading. As expected, the capacity that corresponds to light fluctuations ($m = 10.3$) is larger than the capacity that corresponds to heavy fluctuations ($m = 0.3$). Moreover, this increase is more pronounced for smaller values of $m$. When $\Delta = 0.9$, it is clear from Fig. \ref{oraK}.(a) that the capacity that corresponds to high power of the dominant waves ($K = 10$) is larger than the capacity that corresponds to high power of the scattered waves ($K = 1$). For $\Delta = 1$ in Fig. \ref{oraK}.(b), however, the capacity increases as $K$ decreases, which is consistent with \cite{Rao2015MGF}.}

The simulation, exact and high-SNR approximation BER curves based on (\ref{Pberftr}) and (\ref{PberftrH}) are depicted in Figs. \ref{berm}-\ref{berK}, respectively. We consider the BPSK modulation with modulation parameters $\alpha  = 1$ and $\beta  = 0.5$. Figs. \ref{berm}-\ref{berK} indicate that the analytical expressions coincide with the high-SNR approximation results when average SNR is high. It is clear from Fig. \ref{berm} that a large value of $m$ diminishes the effect of channel fluctuations, thereby delivering a smaller BER. We also note that the BER is a decreasing function of the parameter $K$. With the increase of $K$, the variance ${{\sigma ^2}}$ of the diffuse components decreases under the same average SNR, which results the BER for the case of $K = 25$ is lower than that of $K = 10$.

\section{Conclusion}\label{se:conclusion}
In this paper, we derive new expressions for the PDF and CDF of the instantaneous SNR of the FTR fading channel by using a mixture of gamma distributions. Further, the analytical expression of capacity has been derived. We find that increasing the values of channel parameters $m$ and/or $K$ both help overcome the effects of fading. Moreover, we derived exact and asymptotic expressions of the BER for binary modulation schemes to get better insight into the implications of the model parameters on the BER performance. Our derived results extend the knowledge of the newly proposed FTR fading model, which shows its promising validation for the performance analysis of future wireless systems.

\section*{Appendix}
The PDF of the TWDP fading model can be expressed by a mixture of gamma distributions as \cite[Eq. (6)]{Ermolova2016Capacity}
\begin{align}\label{TWDP}
  {f_{{\text{TWDP}}}}(x) &= \exp ( - K)\sum\limits_{j = 0}^\infty  {\frac{{{K^j}}}{{j!}}} {f_G}\left( {x;j + 1,2{\sigma ^2}} \right) \hfill \notag\\
   &\times \underbrace {\sum\limits_{k = 0}^j {\left( {\begin{array}{*{20}{c}}
  j \\
  k
\end{array}} \right){{\left( {\frac{\Delta }{2}} \right)}^k}\sum\limits_{l = 0}^k {\left( {\begin{array}{*{20}{c}}
  k \\
  l
\end{array}} \right){I_{2l - k}}\left( { - K\Delta } \right)} } }_{{t_j}} \hfill \notag\\
   &= \exp ( - K)\sum\limits_{j = 0}^\infty  {\frac{{{K^j}{t_j}}}{{j!}}} {f_G}\left( {x;j + 1,2{\sigma ^2}} \right), \hfill
\end{align}
where ${I_v}\left(  \cdot  \right)$ is the modified Bessel function of the first kind \cite[Eq. (8.445)]{Gradshteyn1980In} with the \emph{v}-th order, ${f_G}\left( {x;j + 1,2{\sigma ^2}} \right)$ is the PDF of the gamma distribution with the shape parameter (\emph{ j} + 1) and scale parameter ${2{\sigma ^2}}$. In the FTR fading model, $\zeta $ is a unit-mean Gamma distributed random variable with the PDF expression (\ref{fu}). From (\ref{fu}) and (\ref{TWDP}), we can obtain the PDF of the FTR channel as
\begin{align}\label{pdfTWDP2FTR}
 & {f_{{\text{FTR}}}}\left( x \right)= \int_0^\infty  {{f_{{\text{TWDP}}}}\left( {x;uK} \right)} {f_\zeta }\left( u \right)du \hfill \notag\\
&= \frac{{{m^m}}}{{\Gamma \left( m \right)}}\sum\limits_{j = 0}^\infty  {\frac{{{K^j}}}{{j!}}{f_G}\left( {x;j + 1,2{\sigma ^2}} \right)} \sum\limits_{k = 0}^j {\left( {\begin{array}{*{20}{c}}
  j \\
  k
\end{array}} \right){{\left( {\frac{\Delta }{2}} \right)}^k}}  \hfill \notag\\
&\times \sum\limits_{l = 0}^k {\left( {\begin{array}{*{20}{c}}
  k \\
  l
\end{array}} \right)\underbrace {\int_0^\infty  {{u^{j + m - 1}}{e^{ - \left( {m + K} \right)u}}{I_{2l - k}}\left( { - uK\Delta } \right)du} }_{{s_k}}}.
\end{align}
With the help of \cite[Eq. (2.15.3.2)]{prudnikov1986integrals}, we can derive the last term ${{s_k}}$ as
\begin{align}\label{sk}
  {\text{ }}{s_k} &= {e^{\frac{{\pi (2l - k)i}}{2}}}\Gamma \left( {j + m + 2l - k} \right) {\left( {{{(m + K)}^2} - {{(K\Delta )}^2}} \right)^{\frac{{ - (j + m)}}{2}}}  \notag\\
&\times P_{j + m - 1}^{ k - 2l}\left(\frac{{(m + K)}}{{\sqrt {{{(m + K)}^2} - {{(K\Delta )}^2}} }}\right) .
\end{align}
Substituting (\ref{sk}) into (\ref{pdfTWDP2FTR}), we can obtain (\ref{PDF}).

Recall that the definition of the CDF is given by ${F_\gamma }\left( x \right) = \Pr \left( {\gamma  \leqslant x} \right)$, where $\Pr \left(  \cdot  \right)$ represents the probability. With the help of \cite[Eq. (3.351.1)]{Gradshteyn1980In} and ${{f_G}\left( {x;j + 1,2{\sigma ^2}} \right)}$, the CDF expression can be derived as
\begin{align}\label{fg2Fg}
  {F_G}\left( {x;j + 1,2{\sigma ^2}} \right) &= \int_0^t {\frac{{{x^j}}}{{\Gamma \left( {j + 1} \right){{\left( {2{\sigma ^2}} \right)}^{j + 1}}}}\exp \left( { - \frac{x}{{2{\sigma ^2}}}} \right)dt}  \hfill \notag \\
&= \frac{1}{{\Gamma \left( {j + 1} \right)}}\gamma \left( {j + 1,\frac{x}{{2{\sigma ^2}}}} \right). \hfill
\end{align}

Utilizing (\ref{fg2Fg}), we can finish the proof by presenting the CDF of FTR fading model for arbitrary $m > 0$ as (\ref{CDF}).

\bibliographystyle{IEEEtran}
\bibliography{IEEEabrv,Ref}

\end{document}